\documentclass[a4paper]{jpconf}
\usepackage{graphicx}
\usepackage{subfig}
\usepackage{glas}
\begin{document}
\begin{titlepage}{GLAS-PPE/2015-06}{29$^{\underline{\rm{th}}}$ October 2015}
\title{Extending DIRAC File Management with Erasure-Coding for efficient storage.}

\author{Samuel Cadellin Skipsey, Paulin Todev, David Britton, David Crooks and Gareth Roy
\\
School of Physics and Astronomy, Kelvin Building, University of Glasgow, Glasgow, G12 8QQ}

\begin{abstract}The state of the art in Grid style data management is to achieve increased resilience of data via multiple complete replicas of data files across multiple storage endpoints. While this is effective, it is not the most space-efficient approach to resilience, especially when the reliability of individual storage endpoints is sufficiently high that only a few will be inactive at any point in time.
We report on work performed as part of GridPP\cite{GridPP}, extending the Dirac File Catalogue and file management interface to allow the placement of erasure-coded files: each file distributed as N identically-sized chunks of data striped across a vector of storage endpoints, encoded such that any M chunks can be lost and the original file can be reconstructed.
The tools developed are transparent to the user, and, as well as allowing up and downloading of data to Grid storage, also provide the possibility of parallelising access across all of the distributed chunks at once, improving data transfer and IO performance. We expect this approach to be of most interest to smaller VOs, who have tighter bounds on the storage available to them, but larger (WLCG) VOs may be interested as their total data increases during Run 2.
We provide an analysis of the costs and benefits of the approach, along with future development and implementation plans in this area. In general, overheads for multiple file transfers provide the largest issue for competitiveness of this approach at present.

\vspace{0.5cm}
\begin{center}
{\em 21st International Conference on Computing for High Energy and Nuclear Physics (CHEP2015)}\\
{\em Okinawa, Japan}
\end{center}
\end{abstract}

\newpage
\end{titlepage}

\section{Introduction}

WLCG\cite{wlcg} VOs have been distributing data across geographical and administrative boundaries since the very start of the project. Distributed data is beneficial for both performance and reliability  - having data in many places means that it is also near to more compute and other resources, but also means that you can afford to lose access to one or more of those places without losing the data itself.
However, to our knowledge, no WLCG experiment data model has ever broken with the orthodoxy that geographical data distribution implies integer replication of data, one full copy per site.
This is with stark contrast to the internal mechanisms used at the server level, where RAID5\cite{RAID} or 6 stripes and parity-checks individual files. At the data centre level, mature distributed storage systems such as HDFS\cite{hdfs,hdfsraid} and Ceph\cite{ceph} support striping and parity-resiliency of files across servers.
Outside of WLCG, there are successful implementations of the same mechanisms on a globally distributed scale - the DDN WOS\cite{wos} system is one such example, but Tahoe-LAFS\cite{LAFS,LAFS2} provides an analogous model from a diametrically opposed political philosophy.

\subsection{Erasure codes}
Erasure codes (EC) are a general class of forward error correction codes\cite{ec1}, designed to enable the recovery of the content of a message after a subset of the symbols comprising it are lost. In general, most implementations are based on Reed-Solomon\cite{ec2} codes, which have the advantage that they are widely used, implemented and optimised. Alternative, more modern, codes such as Turbo\cite{turbo} codes and LDPC codes provide better performance when transmitting data over a channel, but achieve this via asymptotically amortised overheads as the stream length grows. Similarly, Raptor\cite{raptor} codes (and other fountain codes) provide a probabilistic guarantee of stream recovery with any given number of symbols received. By contrast, Reed-Solomon codes work in terms of a rigid decomposition of the input data into a fixed number of symbols, or chunks, and a fixed number of additional coding symbols (each of which guarantees to be sufficient to recover from loss of any one of the original symbols) which is better suited to data storage across an integer number of endpoints.

The original design of the Reed-Solomon algorithm was based on oversampling of a polynomial. Treating the N chunks of input data as samples to fit an order N polynomial, we can take any additional M samples (the coding chunks), such that any subset of at least N chunks of the new data stream will reconstruct the same polynomial (and hence allow reconstruction of the original N samples). A strict implementation using this algorithm is extremely slow at recovering lost data chunks, and at detecting errors in any chunks in the sequence, and so more modern implementations use a Fourier domain approach based on generating coefficients of a polynomial, rather than sample values\cite{ec3, ec4}.

One of the key advantages of modern Reed-Solomon implementations is that they allow effectively arbitrary selection of the number of chunks in which to section a piece of data, and the number of additional `coding chunks' generated to provide resilience. That is, rather than being limited to specific optimised levels of resilience (such as RAID6, which provides for up to two chunks of data being lost per stripe), we can select any rational value for our resilience and replication levels within reason. Recent suggestions by the ATLAS VO that storage pressures may cause them to maintain less than 2 replicas of data at Tier-2s leave them with no other level of replication but a single copy. With a rational value of `replication' provided by erasure coding, they could tailor their resilience to a finer degree, while still guarding against loss of any single site. 
(In fact, as more than 90\% of SEs are available at any one time, it seems that replicating data twice may be a significant overcommitment to resilience, given the probability of both endpoints for a file being down at once.)

\section{Implementation of a shim for Erasure Coding}

Given the preceding section, the authors felt that the time was long due to attempt to introduce geographically distributed erasure coding techniques to WLCG computing infrastructure. While a more complete implementation would be a deeply integrated component of standard tools, allowing direct remote file access across distributed chunks, the topic of this paper is the initial, proof-of-concept implementation. The principle here is to test how easily we can introduce distributed erasure coding of files without deeply modifying existing tools - the EC layer is simply a shim on top of existing data management.

When originally planned, this project would have provided a shim based around the LFC. However, LFCs are a strongly deprecated component for the large WLCG experiments, and so we made the decision to change to a different data management architecture. In this case we use the DIRAC platform for job submission.

\subsection{DIRAC File Catalog}
DIRAC\cite{DIRAC} is a job and data management suite developed principally by the LHCb experiment.  While originally developed solely for the LHCb VO, the project is also in use by other, small, VOs.  For example, and most importantly to the authors, effort is being expended at Imperial College\cite{ImperialDIRAC} to implement a multiVO DIRAC service to serve our many non-HEP and other small VOs in the UK. 
As part of the data management component, the DIRAC system provides a file catalogue layer (Dirac File Catalogue, DFC), which allows for arbitrary metadata to be applied to individual records (as key-value pairs).

\subsection{ZFEC}
As the DIRAC client and server are written in Python, and provide a Python API, it was advantageous for us to use a similarly Python-compatible library for the EC component of the shim. This requirement steered us towards the zfec\cite{zfec} library, an open-source implementation of Reed-Solomon erasure codes originally developed as a component of the Taho-LAFS filesystem. Whilst the kernel of zfec's encoder is written in C, it provides a fully featured Python API, which is easy to interface with. 
The other obvious alternative Reed-Solomon implementation is the jerasure\cite{jerasure} library. However, at the time at which work was progressing, the author of this library was forced to remove it from public availability due to a patent challenge from the US company StreamScale\cite{jerasure2}. While the code is still used in several derived products (including as the erasure coding library for the Ceph distributed object store), we felt that the legal situation made it difficult for us to justify use of the library at that time.

\subsection{Overlay design}
In the interest of keeping a proof-of-concept design simple, our shim treats grid storage elements essentially as data archives. Files are assumed to be retrieved from the Grid to allow for local file access, so we do not need to support direct IO against the erasure coded data itself. While this mode of data access was common only a few years ago on Grid compute environments, it is increasingly the case that direct IO is used against even remote files by jobs at a site. Implementing direct IO against an encoded file is significantly more difficult, however, and is not supported directly by the DIRAC file catalogue API, so we do not attempt it for this work.

With this restriction, our shim simply performs the erasure coding on the local machine, generating separate temporary files for each chunk and the identically-sized coding chunks. 

On the DIRAC side, we wrap calls to the DFC API, creating a directory in the DFC namespace with the filename requested by the user. Logically, we locate the individual chunks within that directory, with names coded with the standard zfec extensions for chunks (encoding the ordinal number of the chunk in the coding vector, and the total number of chunks and coding chunks expected). As an additional check, and to improve efficiency, we also add metadata to the DFC entries (and directory entry), encoding the total number of chunks (as `TOTAL') and total number of non-coding chunks (as `SPLIT'), as well as some versioning information in case of format changes.

\begin{figure}[h] 
   \begin{minipage}{0.47\linewidth}
   \includegraphics[width=\linewidth]{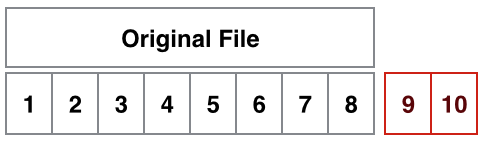} %
    \caption*{Diagram of file splitting for 8 chunks plus 2 additional coding chunks, via (Reed-Solomon) erasure coding.}
   \end{minipage}\hspace{2pc}
 \begin{minipage}{0.47\linewidth}
   \includegraphics[width=\linewidth]{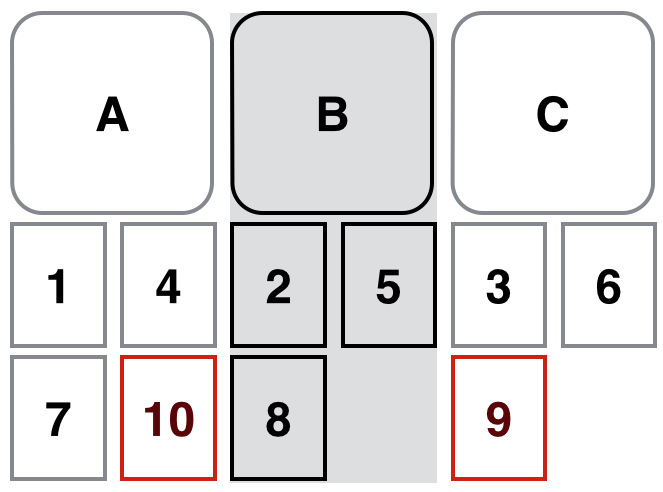} %
    \caption*{ Sample layout for a file split as 8 chunks plus 2 coding chunks (10 chunks overall), distributed across a vector of 3 SEs (A to C)}
   \end{minipage}
   \caption{\label{layout} Layout scheme for shim}
\end{figure}


In order to distribute chunks most effectively, we retrieve a vector of all of the $s$ Storage Element (SE) endpoints supporting the User's VO. Placement is performed as a round-robin loop over this vector, such that chunk 1 is transferred to the first SE endpoint in the vector, and chunk $n$ to the $n$ $mod$ $s$th endpoint. While simple to implement, round-robin data placement suffers from some issues, not least that the first endpoints in the vector will tend to get more chunks over time (assuming that the endpoint vector returned from the DFC is always ordered the same way), see figure \cite{layout}. Only in the case where the number of chunks plus coding chunks is a multiple of the available endpoints will all endpoints receive an equal distribution. Additionally, we do not have any way to factor in geographical distribution in our placement. It is likely that a mature placement algorithm would be best targeted at distribution preferentially across SEs in a geographical region, rather than across the entire world, regardless of network distance.

File retrieval follows the reverse path - the tool retrieves the list of all files in the relevantly named directory in the DFC namespace, and then makes local copies of them via the DIRAC file transfer API. The final file is reconstructed from the on-disk copies of the chunks. As an optimisation, we stop getting chunks as soon as we have enough to reconstruct the file (that is, the same number of chunks as in the original file, disregarding coding chunks).

\subsection{Parallelism}
The above algorithm is functional, but suffers from the disadvantage of being purely serial in operation. As the overhead of erasure coding itself scales with file size (and number of chunks and coding chunks), the serial version of the tool has significant performance deficits compared to a classical file transfer tool. 
Given that the file transfer operations themselves are independent, we can parallelise that portion of get and put operations over a number of threads. 
This is simply achieved by a work pool model at present, where a user-defined set of worker threads are created, and consume file transfer operations until enough chunks have been fetched in total. This has the advantage that a single slow transfer does not necessarily limit the rate at which the file can be retrieved, as the remaining threads can consume the remaining transfer operations at the same time. In the limit where the number of threads is equal to the number of chunks, we essentially select the N fastest chunks out of the total stripe, retrieving the file as fast as the network allows.

\section{Results}

Due to limitations on the supported operating systems for the lcg\_utils packages, development and testing of the shim were performed on a Scientific Linux 6.5 guest virtual machine, hosted on VirtualBox on Windows 8. The well known issues with virtual machine performance compared to bare metal have not been addressed when comparing the measured IO performance in these benchmarks, so they should be treated as purely relative measures of performance. Performance was tested on both single and dual-core systems in order to estimate the effect of contention between threads, but the difference between the two systems was small ($< 5$\%) in almost all cases, so only the single-core results are reproduced here.

\begin{figure}[h] 
   \begin{minipage}{0.47\linewidth}
   \includegraphics[width=\linewidth]{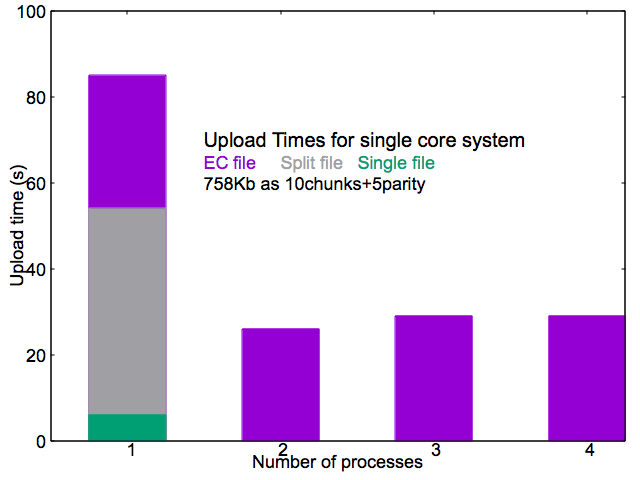} %
    \caption{\label{fig1a} Scaling performance of file upload for a 768kB file encoded as 10 chunks + 5 coding chunks, with increasing parallelism.}
   \end{minipage}\hspace{2pc}
 \begin{minipage}{0.47\linewidth}
   \includegraphics[width=\linewidth]{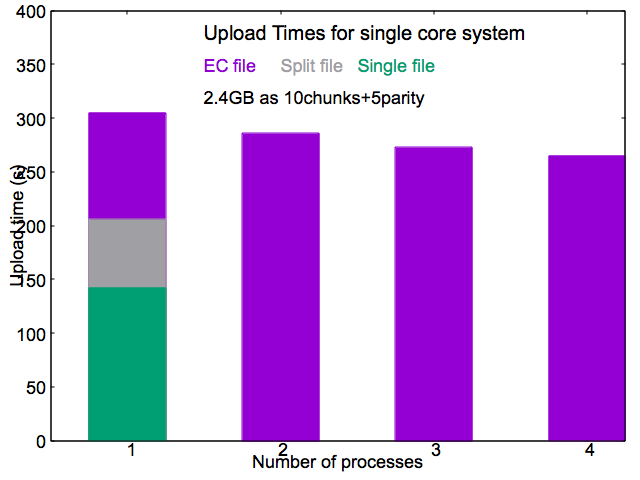} %
    \caption{\label{fig1b} Scaling performance of file upload for a 2.4GB file encoded as 10 chunks + 5 coding chunks. }
   \end{minipage}
\end{figure}

Considering the performance for file upload from the client machine first, we see that for both small (fig \ref{fig1a}) and large (fig \ref{fig1b}) files, parallelism does provide a performance improvement with respect to the single-threaded case. For small files, parallelism improves performance to better than that of the single-threaded case for a file split into 10 pieces with no encoding, but we do not see the same effect for larger files.
This is clearly an Amdahl's Law\cite{amdahl} effect: for small files, file transfer latency dominates the process, and this is distributed over our parallel transfer threads; for larger files, perhaps closer to the size of physics data, the file encoding time is the dominant component, and this is not parallelised in our model.
For smaller files, in fact, the dominant part of the file transfer itself is channel setup and communication, rather than data transfer. From table \ref{table1}, we can see that for a small file split into 10 chunks (without encoding), the time for each chunk transfer is almost identical to that to transfer the whole file. For a larger file, however, the data transfer time itself is dominant, as can be seen by the significantly smaller time it takes to transfer each 1/10 sized chunk. Even in the large file case, there is an overhead to transferring multiple files, which causes the total time to be significantly larger than the single file case. 

\begin{table}[h]
\caption{\label{table1}Comparison of upload times for whole files or files in 10 pieces (with no encoding).}
\begin{center}
\begin{tabular}{|c|c|c|}
\hline
Size&Total time [s]&Average time per file [s]\\
\hline
\hline
1$\times$756 KB&6&6\\
10$\times$75.6 KB&54&5.5\\
\hline
1$\times$2.4 GB&142&142\\
10$\times$243 MB&206&20\\
\hline
\end{tabular}
\end{center}
\end{table}

\begin{figure}[h] 
   \begin{minipage}{0.49\linewidth}
   \includegraphics[width=\linewidth]{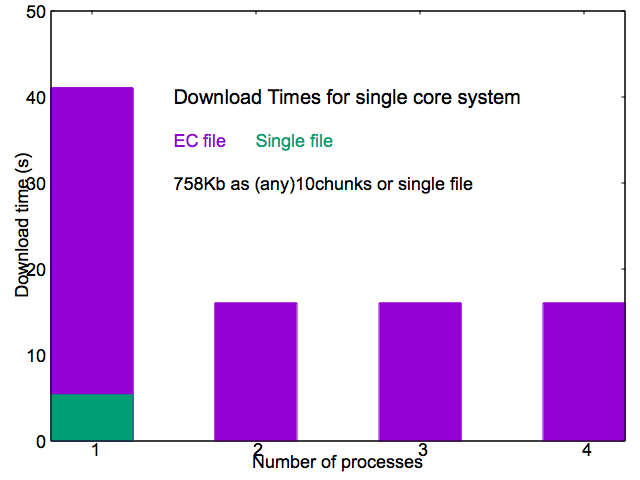} %
    \caption{\label{fig2a} Scaling performance of file download for a 768kB file encoded as 10 chunks + 5 coding chunks, with increasing parallelism.}
   \end{minipage}\hspace{2pc}
 \begin{minipage}{0.49\linewidth}
   \includegraphics[width=\linewidth]{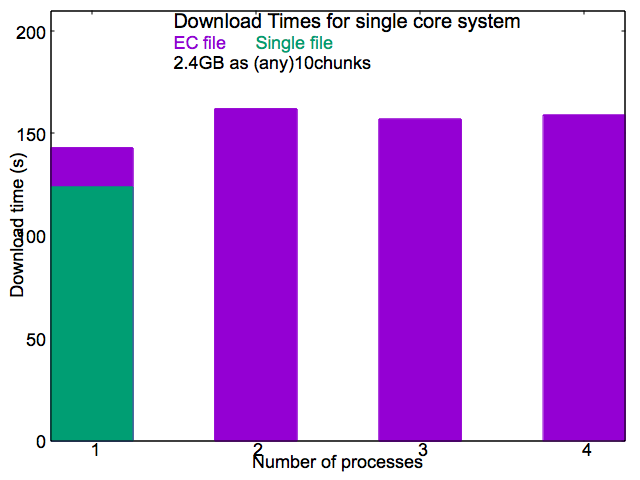} %
    \caption{\label{fig2b} Scaling performance of file download for a 2.4GB file encoded as 10 chunks + 5 coding chunks, with increasing parallelism.}
   \end{minipage}
\end{figure}

We see similar results for file retrieval and decoding. For small files (fig \ref{fig2a}), parallelism significantly improves performance (although not to the level of a single file copy operation on an unencoded file). For larger files (fig \ref{fig2b}), parallelism appears to initially harm performance on our test system, but the overall range of performance is small across all tests. We believe that the limited network bandwidth on our small test system is probably the bottleneck here, and prevents us from drawing strong conclusions concerning the algorithm itself. 
In these cases, performance is affected mostly by the communication overheads in the transfer phase, as file reconstruction requires little overheads if the original data blocks are the first to be retrieved. For this reason, there is no ``grey" split file column in the download graphs.

\section{Further Work}
The code as represented in this document is available online \cite{repos1}. The further work mentioned in this section is localised to a clone of that repository\cite{repos2}, as the original developer has moved on.
At present, as proof of concept code, the shim does not provide a reliable transfer service - any failed transfer for any chunk will cause an upload to fail (and a download to fail, if it causes the number of successfully transferred chunks to fall below the minimum needed for reconstruction). Transfer retries are easy to implement for the serial version, but cause more subtle complexities for parallel transfers (as trying the next SE in the list, for example, disrupts the distribution of chunks across the vector of SEs as a whole). 
Additionally, the initial design of the DFC interface was undertaken without a full appreciation of the global nature of the metadata tag namespace. Subsequently, it has become apparent that our generic names for EC metadata keys are visible to all other users of the Imperial DIRAC's DFC, potentially causing confusion and misuse of the metadata. Later versions of the shim will use unique prefixes for metadata tags to minimise collisions in the tag namespace.

In general, with the move to more network-centric data access and management models for the WLCG VOs, a more useful direction would be to explore the incorporation of similar technologies into federated data storage protocols, such as xrootd. In this case, leveraging the existing federation logic would allow direct IO to encoded data over the network, reducing the transfer overheads for the sparse reads common in some workflows.
Work has begun on such an implementation, which we hope to present at a later conference.

\ack
The authors would like to thank the NA62 VO, in particular Dan Protopopescu, for useful discussions about their data management model, and the Imperial DIRAC team, in particular Janusz Martyniak, for their work on maintaining and developing the infrastructure used for this work.

\end{document}